\documentclass{iopconfser}
\usepackage{graphicx}
\usepackage[utf8]{inputenc}
\usepackage{amsmath,amssymb}
\usepackage{booktabs}
\usepackage{siunitx}
\DeclareSIUnit\pixel{pixel}

\begin{document}
{
\title{Annular x-ray optics offer superior resolution for radiation sensitive samples}

\author{Felix Wittwer$^{1, 2}$, Peter Modregger$^{1, 2}$}

\affil{$^1$Department of Physics, University of Siegen, Siegen, Germany }
\affil{$^2$Center for X-ray and Nano Science, DESY, Hamburg, Germany}

\email{felix.wittwer@desy.de}
}
\begin{abstract}
High-resolution x-ray microscopy requires a high photon flux to measure the signal from weakly scattering samples. 
This exposes samples to high radiation doses, potentially damaging or destroying them through radiation damage.
In this work, we propose the use of annular optics in scanning microscopy as an alternative to full-aperture optics.
Annular optics act as high-pass filters.
Compared to regular optics with the same numerical aperture, annular optics expose the sample to less dose while producing the same signal from small sample features.
Annular optics benefit significantly from the high photon fluxes of the latest x-ray sources to compensate for their overall smaller cross section.
Using numerical simulations, we show that annular optics offer superior optical performance for sample features close to the resolution limit of the optic.
\end{abstract}

\section{Introduction}
Ernst Abbe showed that the resolution of an optical microscope is fundamentally limited by the light wavelength \cite{born2013principles}.
By using wavelengths shorter than one nanometer, x-ray microscopy can potentially resolve samples down to individual atoms.
However, X~rays have drawbacks that limit the achievable resolution \cite{jacobsenXrayMicroscopy2020}:
At x-ray wavelengths, all materials have a refractive index very close to one.
This makes it difficult and expensive to build high-focusing x-ray optics.
Furthermore, X~rays are ionizing radiation that cause radiation damage in the sample.
Especially biological samples can be very sensitive to radiation, tolerating only low radiation doses.
In x-ray microscopy, smaller features tend to produce a weaker signal.
Increasing the photon flux to boost the signal and resolve these features also increases the radiation dose and risks damaging the sample \cite{howellsAssessmentResolutionLimitation2009}.

Here, we present a method to increase the information that can be collected from the sample without increasing the radiation dose.
For scanning microscopy, we propose to increase the signal at high resolutions by using focusing optics with an annular aperture.
Blocking the central part of the aperture creates a high-pass filter \cite{jacobsenXrayMicroscopy2020, goodmanIntroductionFourierOptics2005}.
Annular apertures are already used to increase the resolution in photolithography \cite{kamonPhotolithographySystemUsing1991} and electron microscopy \cite{enyamaMethodImprovingImage2016}.
They were also proposed as a way to increase the signal in x-ray pinhole imaging \cite{ressDemonstrationXrayRingaperture1992}.

\begin{figure}
    \centering
    \includegraphics{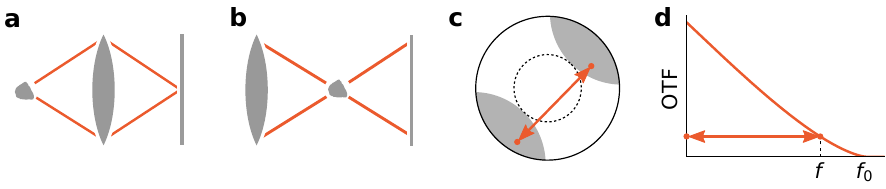}
    \caption{
    (a) In transmission geometry, the lens projects an image of the sample onto the detector.
    (b) In scanning geometry, the lens creates a focal spot that is used to scan the sample. A detector records the generated signal.
    (c) The transmittance and signal for a certain scattering vector (orange arrow) are determined by how many possibilities there are for this vector to fit into the lens aperture (gray area, adapted from \cite{goodmanIntroductionFourierOptics2005}).
    (d) The optical transfer function (OTF) describes the transmission of a lens in relation to the spatial frequency $f$.
    }
    \label{fig:illustration}
\end{figure}
The basic geometries of a transmission and a scanning x-ray microscope are illustrated in Fig.~\ref{fig:illustration}a and b.
In transmission geometry, the lens forms an image by focusing the X rays scattered by the sample.
Abbe's microscopy theory explains that the focusing by the lens causes scattered and unscattered X~rays to interfere and it is this interference which forms the image.
Crucially, this can only occur if both scattered and unscattered X~rays are focused by the lens.
Figure~\ref{fig:illustration}c shows a scattering vector with a random length and orientation.
For large scattering vectors there is only a small aperture section, highlighted in gray, where both the scattered and the unscattered X~rays fit into the lens aperture.
For scattering vectors with the same length but different orientation, the gray region will be rotated. 
The center of the lens however, marked by the dashed circle, will not contribute to the image for any scattering vector of the same length.
The length of a scattering vector is proportional to the spatial frequency of the scattering structure. 
Accordingly, large scattering vectors carry information about small sample features.
Therefore only the rim of the lens is important for high-resolution imaging.

Figure~\ref{fig:illustration}d shows how the size of the effective lens aperture shrinks for large scattering vectors, reducing the transmission of the lens.
Even a perfectly transparent lens will have a weak transmission close to its resolution limit.
For a circular lens aperture, the transmission decreases slightly faster than linear until the resolution limit $f_0$ of the lens.
Because the full aperture contributes to the radiation dose but only the outer ring to the high-resolution signal, we propose to block or remove the central part of the aperture.
This will not affect the image quality at high resolutions but the sample will be exposed to a lower dose.

\section{Background}
Following Goodman \cite{goodmanIntroductionFourierOptics2005}, the image $S(x,y)$ in transmission and scanning microscopy can be described with 
\begin{equation}
    S(x,y) = G(x,y) * P(x,y),
\end{equation}
a convolution of the ground truth distribution $G(x,y)$ with the point-spread function (PSF) $P(x,y)$ of the lens.
According to the convolution theorem, the Fourier transform of $S$ is the product of the Fourier transforms of $G$ and $P$:
\begin{equation}
    \mathcal{F}\{S(x, y)\} = s(f_X, f_Y) = g(f_X, f_Y)\cdot p(f_X, f_Y),
\end{equation}
with spatial frequencies $f_X$ and $f_Y$.
In incoherent imaging, the PSF is the absolute square of the wavefield $A(x,y)$ in the sample plane, which is the inverse Fourier transform of the aperture function~$a(f_X,f_Y)$:
\begin{equation}
    P(x,y) = |A(x,y)|^2 = \left|\mathcal{F}^{-1}\{a(f_X, f_Y)\}\right|^2.
\end{equation}
Accordingly, $p$ is the auto-correlation of the aperture function:
\begin{equation}
    p(f_X, f_Y) = a(f_X, f_Y) * \overline{a(f_X, f_Y)}.
\end{equation}
To better compare different optical systems, $p$ is often normalized to the transmitted intensity. 
The result is known as the optical transfer function (OTF):
\begin{equation}
    \mathit{OTF}(f_X, f_Y) = \frac{p(f_X, f_Y)}{p(0, 0)}. \label{eq:otf_definition}
\end{equation}
If the aperture has a radial symmetry, the OTF depends only on $f = \sqrt{f_X^2 + f_Y^2}$.
Altogether, the microscopy image can be modeled via
\begin{equation}
    S(x,y) = I_0 \cdot G(x,y) * \mathcal{F}^{-1}\{OTF(f_X, f_Y)\}, \label{eq:imaging_equation}
\end{equation}
where $I_0$ is the total photon intensity.
For an ideal circular lens, the aperture function is a binary mask.
Figure~\ref{fig:illustration}d shows the OTF when the sample is in the focal plane of the lens, achieving the highest resolution.
The resolution limit $f_0$ is given by
\begin{equation}
    f_0 = \frac{d}{\lambda z},
\end{equation}
with lens diameter $d$, focal length $z$ and wavelength $\lambda$ \cite{goodmanIntroductionFourierOptics2005}.
The plot highlights a critical aspect of transmission and scanning microscopy: the signal strength decreases for high spatial frequencies, resulting in a weaker signal from smaller features.
For sample features twice the size of the resolution limit, the detected signal is already less than \qty{40}{\percent} of the full signal strength. 
Close to the resolution limit, the signal strength is practically zero.

\section{Numerical simulation}
\begin{figure}
    \centering
    \includegraphics{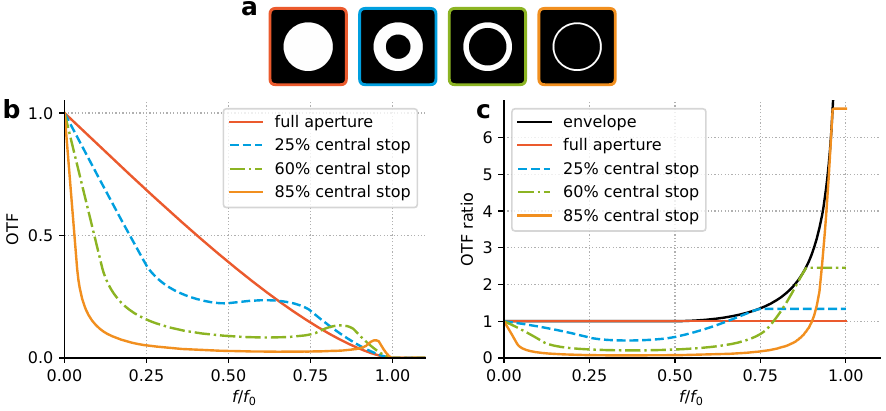}
    \caption{
    (a) Four annular apertures with different central stops: Full circular aperture (red), central stop blocking \qty{25}{\percent} (blue), \qty{60}{\percent} (green), or \qty{85}{\percent} (yellow) of the aperture.
    (b) Optical transfer function (OTF) for each of the four apertures.
    (c) Ratio between each OTF and the OTF of the full aperture.
    The envelope is calculated from all possible central stop sizes.
   }
    \label{fig:otf_plot}
\end{figure}
Annular apertures are advantageous when the sample can only be exposed to a certain radiation dose.
To demonstrate this advantage, we use numerical simulations to model scanning transmission x-ray microscopy (STXM) measurements.
The simulations model a lens with a diameter of \qty{260}{\micro\meter} and a focal length of \qty{205}{\milli\meter} at a wavelength of \qty{0.1}{\nano\meter}, resulting in a resolution limit of \qty{80}{\nano\meter}.
The simulations compare four different scenarios.
The first scenario uses the full aperture, the other three use central stops with diameters of \qty{130}{\micro\meter}, \qty{200}{\micro\meter}, and \qty{240}{\micro\meter}.
Figure~\ref{fig:otf_plot} shows the OTFs for the four scenarios, simulated with (\ref{eq:otf_definition}).
Using a central stop reduces the transmission at low spatial frequencies.
However, because the OTF is normalized to the total intensity, the transmission increases at high spatial frequencies compared to the full aperture.
For the thinnest ring with only \qty{15}{\percent} of the original aperture, the OTF at the highest spatial frequencies is more than six times higher than for the full aperture.
Consequently, a high-resolution scan using this ring aperture requires less than one sixth of the dose of a full aperture scan.

\begin{figure}
    \centering
    \includegraphics{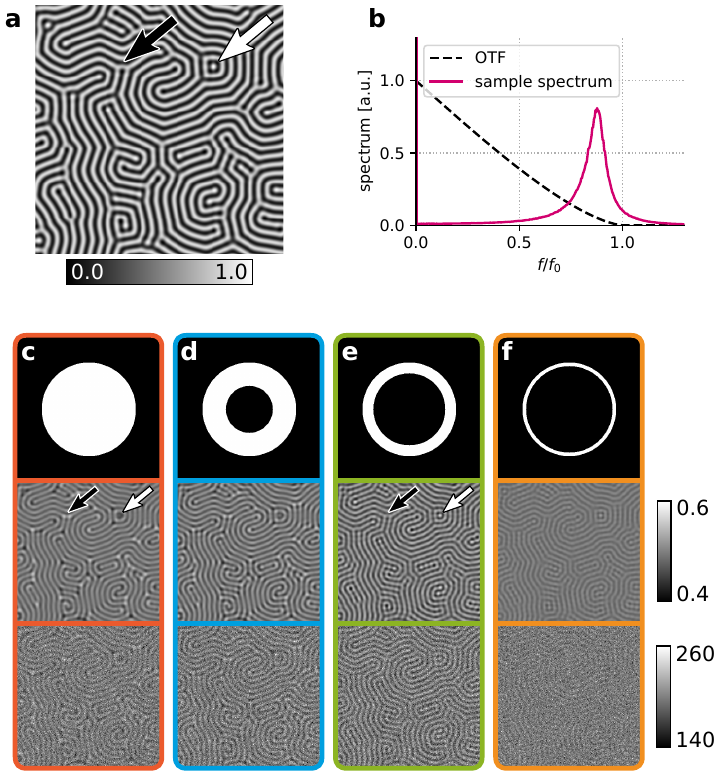}
    \caption{
    (a) A \qtyproduct{2.56 x 2.56}{\micro\meter} section of the Swift-Hohenberg phantom. The black arrow indicates a T-junction of the bight lines, the white arrow indicates an isolated bright spot. (b) Azimuthally averaged spatial frequencies of the phantom. The critical frequency is around $0.9f_0$, corresponding to a spacing of \qty{90}{\nano\meter} per line. The dashed line indicates the OTF of the full aperture lens with a resolution of \qty{80}{\nano\meter}.
    (c-f) Simulated absorption scans of the phantom. Each column shows a different scenario with the central stop blocking (c) \qty{0}{\percent}, (d) \qty{25}{\percent}, (e) \qty{60}{\percent}, (f) \qty{85}{\percent}. The first row shows the lens aperture in each scenario. The middle row shows noise-free measurements. The bottom row shows scans with Poisson noise added equivalent to \num{400} incident photons per scan point.}
    \label{fig:combined_sample}
\end{figure}
The simulations model the scan of a phantom sample generated from the Swift-Hohenberg equation \cite{swiftHydrodynamicFluctuationsConvective1977}.
This equation can describe certain biological patterns, like those arising from reaction-diffusion systems, exhibiting a  characteristic length scale.
The phantom has a size of \qtyproduct{205 x 205}{\micro\meter} and a resolution of \qty{10}{\nano\meter}. 
Figure~\ref{fig:combined_sample}a shows a \qtyproduct{2.56 x 2.56}{\micro\meter} section of the phantom, used in all plots. 
Figure~\ref{fig:combined_sample}b shows the spectrum of the sample.
The low spatial frequencies of the phantom are negligible, indicating that the line pattern has no long-range order.
The characteristic pitch of the pattern lines is \qty{90}{\nano\meter}.

The simulations compare the four scenarios for two cases, a noise-free measurement and a measurement with Poisson noise equivalent to \num{400}~photons per scan point.
The STXM scans were simulated with (\ref{eq:imaging_equation}) and the results are shown in Fig.~\ref{fig:combined_sample}c-f.
All eight simulations show a reduced contrast in comparison to the ground truth image.
This is because the \qty{90}{\nano\meter} line pitch is close to the \qty{80}{\nano\meter} resolution limit of the lens, see Fig.~\ref{fig:combined_sample}b.
In the first scenario, using the full aperture (red case), the lines appear blurred and muted.
There are also notable artifacts in the scan image although the relevant length scales are larger than the resolution limit.
The T-junctions, indicated by the black arrow, are much brighter than the lines itself.
This is an artifact that is not present in the original phantom.
The scan also fails to resolve the isolated bright spot indicated by the white arrow.
In the second scenario, with \qty{25}{\percent} of the aperture blocked (blue case), the result is nearly identical to the first scenario.
The lines appear slightly sharper and the highlights at the T-junctions are less bright, but the isolated spot is not resolved.
In the third scenario, with \qty{60}{\percent} of the aperture blocked (green case), the result is different.
The contrast between the bright and dark lines is improved, the junctions are no longer brighter than the lines and even the isolated spot is clearly resolved.
This remains unchanged even when Poisson noise is added to the measurement. 
In the last scenario, with \qty{85}{\percent} of the aperture blocked (yellow case), the resolution is similar to the previous case.
The lines are sharp, with no part brighter than any other part.
At the same time however, the contrast has decreased dramatically.
Consequently, when Poisson noise is added to the measurement, the structures disappear in the noise.

\begin{table}
    \centering
    \caption{Structural similarity index between ground truth and scan images}
    \begin{tabular}{rcccc}
        \toprule
        central stop size & \qty{0}{\percent} & \qty{25}{\percent} & \qty{60}{\percent} & \qty{85}{\percent} \\
        \midrule
        noise-free & \num{0.126} & \num{0.160} & \num{0.217} & \num{0.086} \\
        Poisson noise & \num{0.123} & \num{0.156} & \num{0.212} & \num{0.084} \\
        \bottomrule
    \end{tabular}
    \label{tab:structural_similarity}
\end{table}
We quantify these results using the structural similarity index measure (SSIM) \cite{zhouwangMeanSquaredError2009} to compare the simulated measurements to the ground truth image.
The results in Table~\ref{tab:structural_similarity} support the visual impression that the third scenario is most similar to the ground truth.
The central stop blocking \qty{60}{\percent} of the aperture gives the optimal result with the best resolution and contrast for this sample phantom.
The difference between the noise-free



\section{Discussion}
Optics with an annular aperture have a smaller cross-section than a full aperture optic with the same diameter and focal length.
For the same incident flux and exposure time, the radiation dose to the sample will be lower.
Crucially, the high-resolution signal remains unchanged.
Alternatively, if the exposure time can be increased or a higher incident flux is available, the sample can be exposed to the same dose as before, but with a boosted high-resolution signal.
This possibility is significant for fourth generation synchrotron radiation sources.
Fourth generation sources are expected to generate two orders of magnitude more photons than the previous generation.
The signal strength in measurements of radiation-sensitive samples will be limited not by the available flux but by the dose that the sample can tolerate.
In this case, the smaller cross-section of annular optics can be compensated by using the full incident photon flux and allows to trade the signal strength at low spatial frequencies for an increased signal at high resolutions.
This works best for samples with a negligible spectrum at low spatial frequencies, like in the numerical simulation.
If the long-range information is important, it is possible to recover it with a deconvolution with the PSF of the annular optic \cite{enyamaMethodImprovingImage2016}.



\section{Conclusion}
Annular aperture optics act as high-pass filters that are only sensitive to high spatial frequencies, suppressing low-resolution features.
Because imaging and scanning geometry are mathematically equivalent, a scanning probe formed from an annular aperture is also most sensitive to high spatial frequencies.
In a scan that aims for the highest resolution, a probe from an annular aperture can achieve the same resolution as a full aperture probe at a lower radiation dose to the sample.

We demonstrated this on a numerical simulation comparing scans of a Swift-Hohenberg phantom using four different apertures.
The best result is achieved with a central stop that blocks \qty{60}{\percent} of the aperture.
Apertures with smaller or no central stop fail to capture all features of the sample, while larger central stops lead to reduced contrast that disappears in the noise.

One drawback of annular apertures is their poor performance at low spatial frequencies.
Semi-transparent central stops might improve this, resulting in a more even OTF across the spatial frequencies.
Central stops have the advantage of being easy to integrate into current optics and allowing a switch between different modes that are optimized for certain feature sizes.
Dedicated optics with an annular aperture might be attractive for simpler manufacturing and because the central area can be used for other instruments, for example a sample holder with a short working distance.

\section*{Acknowledgments}
This research was supported in part through the Maxwell computational resources operated at Deutsches Elektronen-Synchrotron DESY, Hamburg, Germany.

\bibliographystyle{vancouver}
\bibliography{references}

\end{document}